\documentclass[pra,twocolumn,titlepage,nofootinbib,amsmath,showpacs]{revtex4}%
\usepackage{graphicx}%
\newcommand{\eq}[1]{(\ref{#1})}
\newcommand{\alphab}{\mbox{\boldmath$\alpha$}}
\newcommand{\Za}{Z\alpha}
\newcommand{\Zab}{(Z\alpha)}
\newcommand{\calO}[1]{{\cal O}\left( #1 \right)}
\newcommand{\someletter}{\varepsilon}

\begin{document}
\title{Relativistic recoil effects for energy levels in a muonic atom
within a Grotch-type approach: An application to the one-loop electronic
vacuum polarization}
\author{Vladimir G. Ivanov}
\affiliation{Pulkovo Observatory, St.Petersburg, 196140,
Russia}
\author{Evgeny Yu. Korzinin}
\affiliation{D.~I. Mendeleev Institute for Metrology, St.Petersburg,
190005, Russia}
\author{Savely~G.~Karshenboim}
\email{savely.karshenboim@mpq.mpg.de}
\affiliation{Max-Planck-Institut f\"ur Quantenoptik, Garching,
85748, Germany} \affiliation{Pulkovo Observatory, St.Petersburg,
196140, Russia}

\begin{abstract}
We continue our account of relativistic recoil effects
in muonic atoms and present explicitly analytic results at
first order in electron-vacuum-polarization effects. The results are obtained
within a Grotch-type approach based on an effective Dirac equation. Some
expressions are cumbersome and we investigate their asymptotic behavior.
Previously relativistic two-body effects due to the one-loop electron vacuum
polarization were studied by several groups. Our results found here are
consistent with the previous result derived within a Breit-type approach
(including ours) and disagree with a recent attempt to apply a Grotch-type
approach.
\end{abstract}
\maketitle

\newpage

\section{Introduction}

An analytic calculation of relativistic recoil effects in a
hydrogen-like atom to order $(Z\alpha)^4m^2/M$ is possible through
the equation \cite{Gro67}
\begin{equation}\label{eq:gro}
E = m + m_R
\Bigl(f_{C}({Z\alpha})-1\Bigr)-\frac{m_R^2}{2M}\Bigl(f_{C}({Z\alpha})-1\Bigr)^2
\,,
\end{equation}
where $f_{C}({Z\alpha})$ is the dimensionless energy of a
Dirac-Coulomb equation, which is indeed well known (see, e.g.,
\cite{IV}). The corrections are of order
${\cal O}((Z\alpha)^4(m/M)^2m)$ and ${\cal O}((Z\alpha)^5(m/M)m)$.
The terms in $(Z\alpha)^5$ are due to effects of multiphoton exchange. It is even possible
to provide a complete calculation
of the $m/M$ recoil effects for pure Coulomb two-body systems by taking
into account multiphoton exchange contributions exactly in $(Z\alpha)$
\cite{shabaev,yelkhovsky}.

As was shown in our previous paper \cite{I}, one can consider a more general
problem and the result for the energy takes the form
\begin{eqnarray}\label{eq:1:exa}
E &=& m + m_R \Bigl(f_{CN}({Z\alpha},\kappa)-1\Bigr)\nonumber\\
&-&\frac{m_R^2}{2M}\Bigl(f_{CN}({Z\alpha},\kappa)-1\Bigr)^2\nonumber\\
&-&\frac{m_R^2}{2M}\frac{\partial}{\partial \ln\kappa}
\Bigl( f_{CN}({Z\alpha},\kappa)-1\Bigr)^2\nonumber\\
&-&\langle \psi|\left( \frac{V^2}{2M} +
\frac{1}{4M} [V,[\mathbf{p}^2,W]]
  \right) |\psi\rangle\,,
\end{eqnarray}
where $\kappa=Z\alpha m_R/\mu$,
the potential is
\[
V=V_C + V_N
\,,
\]
where in certain sense $V_N\sim \someletter V_C$, $\someletter\ll1$.
Here, $W$ is a specific auxiliary potential, $\psi$
is the wave function of the Dirac problem with the reduced
mass and $f_{CN}({Z\alpha},\kappa)$ is the
dimensionless energy for the potential $V_C+V_N$. The momentum scale
(i.e. the characteristic inverse radius) of $V_N$ is $\mu^{-1}$. For the case of the
Uehling potential the scale parameter is defined as $\mu=m_e$.

In this paper we study a correction to the energy in the first
order of $V_N$, so we can write
\[
f_{CN}({Z\alpha},\kappa)=f_C({Z\alpha})
+f_N({Z\alpha},\kappa)\;,
\]
where $f_{N}({Z\alpha},\kappa)$ is the corresponding
dimensionless correction.

Since we are interested only in terms of order $\someletter(Z\alpha)^4m^2/M$,
we can further simplify this expression
\begin{eqnarray}\label{eq:1:nr}
E &=&m + m_R \Bigl(f_{CN}({Z\alpha},\kappa)-1\Bigr)+\Delta E\nonumber\\
\Delta E&=&-\frac{m_R^2}{2M}\Bigl(f_{CN}({Z\alpha},\kappa)-1\Bigr)^2\nonumber\\
&~&-\frac{m_R^2}{2M}\frac{\partial}{\partial
\ln\kappa}\frac{\bigl(E^{\rm (NR)}(\kappa)\bigr)^2}{m_R^2}\nonumber\\
&~&-\langle \psi_{\rm NR} |  \left( \frac{V^2}{2M} + \frac{1}{4M}
[V,[\mathbf{p}^2,W]]
  \right) | \psi_{\rm NR} \rangle\,,
\end{eqnarray}
where it is sufficient to apply the nonrelativistic approximation to the energy in
the term with derivative
\begin{equation}\label{enr}
(f_{CN}({Z\alpha},\kappa)-1)\bigg|_{\rm
nonrel}=\frac{E^{\rm (NR)}(\kappa)}{m_R}\;,
\end{equation}
as well as to the wave function in the last term.

It is remarkable that in certain respects the relativistic recoil
correction beyond the Dirac equation with the reduced mass, $\Delta
E$, is simpler than the solution of the Dirac equation. To
order $\someletter(Z\alpha)^4m^2/M$ it requires
only nonrelativistic evaluation. In particular, the leading recoil
correction, being expressed in pure nonrelativistic terms, does not
depend on the total angular momentum, $j$, but only on the angular
momentum $l$. That means that this correction may contribute to the
Lamb splitting (a difference between states with the same $j$, but
different $l$, such as the $2p_{1/2}-2s_{1/2}$ difference), but not
to the fine-structure interval (a difference between states with the
same $l$, but different $j$, such as the $2p_{3/2}-2p_{1/2}$
difference).

\begin{figure}[htpb]
\begin{center}
 \resizebox{0.14\textwidth}{!}
{\includegraphics{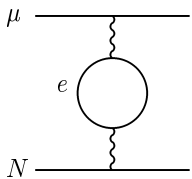}}
\end{center}
\caption{One-photon-exchange diagram for the eVP contributions.  It
is responsible for the the Uehling-potential corrections to orders
$\alpha(Z\alpha)^2m$ and $\alpha(Z\alpha)^4m$. \label{f:u1gamma}}
\end{figure}

To validate applicability of this expression for the
electron-vacuum-polarization (eVP) effects we should prove that the
relativistic recoil effects can be reduced to the evaluation of the
one-photon exchange (see Fig.~\ref{f:u1gamma}) and present explicit
expressions for related contributions to $V$ and $W$.

Apparently, the correction to the potential is the Uehling potential,
which can be presented, e.g., in the form \cite{Schwinger}
\begin{equation}\label{e:uehling}
V_U(r) = -\frac{\alpha\Zab}{\pi} \int_0^1
  dv \, \rho_e(v) \frac{e^{-\lambda r}}{r}
  \,,
\end{equation}
where
\begin{eqnarray}
\lambda&=&\frac{2m_e}{\sqrt{1-v^2}}
  \,,\nonumber\\
  \rho_e(v)&=&\frac{v^2(1-v^2/3)}{1-v^2}\,.
\end{eqnarray}
The factor $\alpha/\pi$ plays a role of the parameter $\someletter$ in our
general consideration \cite{I} and $\mu=m_e$. Meanwhile, a
construction of $W$, which is to be preceded by a choice of an
appropriate gauge is not trivial (see the discussion in
\cite{pra}).

In principle, the correction $\Delta E$ can be treated
relativistically without any nonrelativistic reduction of the
energy and the wave function. However, the higher-order effects
which are incorporated in this case are smaller than possible
effects of two-photon corrections. In particular, for the eVP
contributions the higher-order relativistic-recoil contributions to
$\Delta E$ are of order $\alpha(Z\alpha)^6m^2/M$, while the
two-photon-exchange diagrams contribute to order
$\alpha(Z\alpha)^5m^2/M$.

In the following sections we briefly reproduce this discussion and
present appropriate results for a relativistic and nonrelativistic
correction to the energy due to a Dirac equation with a potential
which accounts for the eVP effects. Using them, we present an analytic
expression for eVP relativistic recoil corrections in the general case
as well as for most interesting particular cases, such as circular
and low-lying states. For both kinds of the states we also derive
their large-kappa asymptotics. In conclusion we discuss a comparison
with an alternative technique for relativistic recoil corrections
based on the Breit-type equation.

\section{Grotch-type expression for the eVP corrections in first order in $\alpha$\label{sec:grotch}}


A choice of gauge for the photon propagator is crucial for the
explicit presentation of the one-photon contribution and for the value
of the two-photon contribution. We have already discussed that in
part in \cite{I} and in detail in \cite{pra}.

Indeed, due to the gauge invariance of quantum electrodynamics, any
final complete result for any physical calculation does not depend
on the choice of the gauge. However, the technical origin of
different contributions to such a final result may be different in
different gauges. In particular, the physical result for a
relativistic recoil correction to order $\alpha(Z\alpha)^4m^2/M$
does not necessarily come only from the static part of the
one-photon-exchange term.

\begin{figure}[htbp]
\begin{center}
  \resizebox{0.40\textwidth}{!} {
  \includegraphics{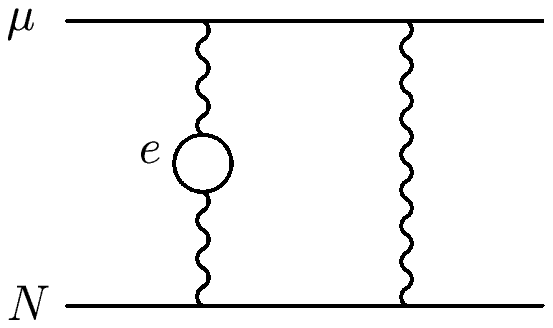} \resizebox{0.03\textwidth}{!} {\ } \includegraphics{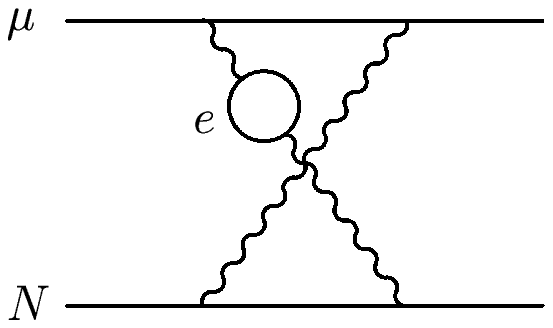}
  }
\end{center}
\caption{Two-photon-exchange diagrams for the eVP contribution.
Subtraction terms and reducible contributions are omitted. In
certain gauges the two-photon-exchange effects contribute to order
$\alpha(Z\alpha)^4m^2/M$. \label{f:u2gamma}}
\end{figure}

Following \cite{pra}, we use the Coulomb gauge for the free photon
propagator, while the eVP correction to the propagator takes the
form
\begin{eqnarray}\label{choice1}
D_{00}^e&=&-\frac{\alpha}{\pi}\int_0^1{dv}\rho_e \frac{1}{({\mathbf k}^2+\lambda^2)}\;, \nonumber\\
D_{i0}^e&=&0\;,\nonumber\\
D_{ij}^e&=&-\frac{\alpha}{\pi}\,\int_0^1{dv}\rho_e \frac{1}{(k^2-\lambda^2)}\nonumber\\
&~& \times \left(\delta_{ij}-\frac{k_ik_j}{({\mathbf
k}^2+\lambda^2)}\right)\;.
\end{eqnarray}
We note that similarly to the Coulomb gauge the $D_{00}$ component
of the photon propagator does not depend on the energy
transfer and $D_{i0}=0$. This choice is sufficient for vanishing
$\alpha(Z\alpha)^4m^2/M$ contributions from two-photon exchanges
(see Fig.~\ref{f:u2gamma}) and thus the problem of calculations of
relativistic recoil effects at this order is reduced to
consideration of the one-photon-exchange diagrams (see
Fig.~\ref{f:u1gamma}).

The Grotch-type calculations of the eVP contribution of order
$\alpha(Z\alpha)^4m^3/M^2$ were considered some time ago
\cite{borie82, borie05,borie11} (for earlier evaluations see \cite{friar73,barrett73}).
However, the gauge was not appropriate and two-photon-exchange corrections
should be added. Those corrections were missed in \cite{borie82,
borie05,borie11}, which produces a discrepancy between the
Breit-type calculation \cite{jentschura} and the Grotch-type ones.
The situation was clarified in \cite{pra}.


Once an appropriate gauge is chosen, we can restrict our
consideration to the static part of the one-photon-exchange
contribution, including the vacuum polarization. Treating the
nucleus nonrelativistically we arrive at the same equation as for
the free one-photon-exchange%
\begin{eqnarray}\label{Dirac}
  &&\biggl(
    \alphab \cdot {\mathbf p}
    + \beta m
    + \frac{\mathbf{p}^2}{2M}
    + V
    + \frac{1}{2M} \left\{ \alphab \cdot {\mathbf p}, V
    \right\}\nonumber\\
    &~& +\frac{1}{4M}
      \left[
         \alphab \cdot {\mathbf p},
         [\mathbf{p}^2, W]
      \right]
  \biggr) \psi(r)
  =
  E \psi(r)
  \,.
\end{eqnarray}
where, however, the effective potentials $W$ and $V$ include eVP
effects and, in particular,
\begin{eqnarray}
  V(r)&=&V_C(r)+V_U(r)\;,\nonumber\\
  V_U({\mathbf k}) &=& -4\alpha\Zab \int_0^1
  dv \, \frac{\rho_e(v)}{{\mathbf k}^2+\lambda^2}\;,\nonumber\\
\end{eqnarray}
and
\begin{eqnarray}\label{wu1}
  W&=&W_C+W_U\,,\nonumber\\
  W_U({\mathbf k})&=&
  -2\frac{V_U({\mathbf k})}{{\mathbf k}^2+\lambda^2}\nonumber\\
  &=&  8\alpha\Zab \int_0^1
  dv \,  \frac{\rho_e(v)}{({\mathbf k}^2+\lambda^2)^2}
  \,,\nonumber\\
  W_U(r) &=& \frac{\alpha\Zab}{\pi} \int_0^1
  dv \, \rho_e(v) \frac{e^{-\lambda r}}{\lambda}
  \,,
\end{eqnarray}
and, as given in \cite{Gro67,I}
\begin{eqnarray}\label{wc}
  W_C(r) &=& -\Za r
  \,,\nonumber\\
  W_C({\mathbf k}) &=& \frac{8\pi\Za}{{\mathbf k}^4}
  \,.
\end{eqnarray}


With an expression for $W$ in hand, we can rewrite the
addition to the Hamiltonian, which in the leading order of $\alpha$,
takes the form
\begin{eqnarray}\label{u:dh}
  \delta H =
  &-&\underbrace{\left( \frac{V_C^2}{2M} + \frac{1}{4M} [V_C,[\mathbf{p}^2,W_C]] \right)}_{=0}\nonumber\\
    &-& \frac{V_U V_C}{M}
    - \frac{1}{4M} [V_C, [\mathbf{p}^2, W_U] ]\nonumber\\
    &-& \frac{1}{4M} [V_U, [\mathbf{p}^2, W_C] ]
  \,.
\end{eqnarray}
The expression turns out to be equal to zero for the pure Coulomb case.


For the case of $V_N=V_U$ and $f_N=f_U$ Eq.~(\ref{eq:1:nr}) leads
to the expression for the following correction of the first
order in $\alpha$
\begin{eqnarray}\label{eq:1:all}
E_U &=&m_R f_{U}({Z\alpha},\kappa)\nonumber\\
&-&\frac{m_R^2}{M}\Bigl(f_{C}({Z\alpha})-1\Bigr)f_{U}({Z\alpha},\kappa)\nonumber\\
&-&\frac{m^2_R}{M} \Bigl(f_C({Z\alpha})-1\Bigr) \frac{\partial}{\partial
\ln\kappa}f_U({Z\alpha},\kappa)\nonumber\\
&-&\langle \psi |  \Bigl(
 \frac{V_U V_C}{M}
    + \frac{1}{4M} [V_C, [\mathbf{p}^2, W_U] ]
\nonumber\\
 && ~~ + \frac{1}{4M} [V_U, [\mathbf{p}^2, W_C] ]\Bigr) | \psi
 \rangle\,.
\end{eqnarray}
The expression includes a nonrelativistic term of order
$\alpha(Z\alpha)^2m$ (exact in $m/M$), a pure relativistic one
$\alpha(Z\alpha)^4m$ and a relativistic recoil correction to order
$\alpha(Z\alpha)^4m^2/M$. It also contains a higher-order
$\alpha(Z\alpha)^6m$ non-recoil term which may be numerically
comparable to $\alpha(Z\alpha)^4m^2/M$ in a certain range of $Z$.

This expression may be approached analytically or by numerical
means. Below we express it in terms of certain base integrals which
we evaluate in closed analytic form following
\cite{rel1,rel2,uehl_cl,uehl_an,soto,quacl,3f2}.


The required expressions for the dimensionless $f_{C}({Z\alpha})$
energy and Dirac-Coulomb wave functions $\psi$ are summarized in the
Appendix of our previous paper \cite{I}.

To obtain final results we have to find some derivatives and it is
important to have an expression for $f_{U}({Z\alpha},\kappa)$ (the
dimensionless Uehling corrections to the energy levels of the
Dirac-Coulomb equation with the reduced mass) in a form suitable for
differentiation and it is also available. Not only analytic
expressions for relativistic \cite{rel1,rel2} (see Sec.~\ref{s:du})
and nonrelativistic \cite{uehl_cl,rel2,uehl_an,soto} (see
Sec.~\ref{s:du}) corrections are known, but also their various
asymptotics \cite{rel1,rel2,uehl_an,soto,quacl,3f2}.

For the analytical differentiation one can take into account that
\[
  \frac{\partial }{\partial \ln\kappa}f(\kappa/n)
  =
  \frac{\partial }{\partial \ln\kappa_n}f(\kappa_n)
\,,
\]
where $\kappa_n=\kappa/n$ is a combination which naturally appears in
various analytic expressions (see Sec.~\ref{s:du}).
For numerical evaluations of the derivative a more useful relation
is
\[
  \frac{\partial}{\partial \ln\kappa}f(\kappa)
  =
  -\frac{\partial }{\partial \ln m_e}f(\kappa)
\,.
\]

\section{The Uehling-potential correction to the energy levels of the
Dirac-Coulomb equation\label{s:du}}

The Uehling correction to the energy of the Dirac-Coulomb+Uehling
problem was addressed in \cite{rel1} for circular states and later
was generalized for an arbitrary state \cite{rel2}. The result reads as
\begin{eqnarray}\label{DeltaEU}
  f_U(\Za,\kappa)
  &=&
  -\frac{\alpha\Zab}{\pi m_R} \nonumber\\
  &&\times
  \int_0^1  dv \, \rho_e(v)
  \langle \psi | \frac{e^{-\lambda r}}{r} | \psi \rangle
  \nonumber\\
  &=&
  \frac{\alpha\Zab^2}{\pi n^2} F_{nl_j}(\kappa'_{n})
  \,,
\end{eqnarray}
where
\begin{eqnarray}
  \kappa'_{n}&=& \frac{\eta m_R}{m_e}\,,\\
\eta &=&\frac{Z\alpha}{\sqrt{(n_r+\zeta)^2+\Zab^2}} \,,\nonumber\\
\zeta &=& \sqrt{\nu^2 -(Z\alpha)^2} \,,\nonumber\\
\nu &=& (-1)^{j+l+1/2} (j+1/2) \,,\nonumber\\
 n_r &=& n - |\nu|\,.\nonumber
\end{eqnarray}

The parameter $\kappa'_n$ is different from
$\kappa_n=\kappa/n=Z\alpha m/n m_e$. However, in a nonrelativistic
approximation
\begin{eqnarray}
  \kappa'_{n}&=& \frac{Z\alpha m_R}{m_e\sqrt{(n_r+\zeta)^2+\Zab^2}}\nonumber\\\
&\simeq& \frac{Z\alpha m_R}{m_e\sqrt{(n_r+|\nu|)^2+\Zab^2}}\nonumber\\\
&=& \frac{Z\alpha m_R}{m_e n}=\kappa_n\;.
\end{eqnarray}

The function $F_{nl_j}(\kappa)$ can be expressed either in terms of a
one-dimensional integral over elementary functions or in terms of a
hypergeometric function $_3F_2$ \cite{rel1,rel2}. Indeed, the
correction $f_U$ can be computed numerically for any desired state.
However, because of a required expansion and various further
considerations, such as examination of the asymptotic behavior, we
prefer here analytic or semi-analytic results.

In particular, as it was found in \cite{rel2}
\begin{equation}\label{Fnlj}
  F_{nl_j}(\kappa)
  =
  -\sum_{i,k=0}^{n_r} B_{ik} \, K_{2,2\zeta+i+k}(\kappa) 
  \,,
\end{equation}
where
\begin{eqnarray}
  B_{ik}
  &=&
  \left(\frac{\eta n}{\Za}\right)^2 
  \,
  \frac{(-1)^{i+k}(n_r)!}{i!(n_r-i)!k!(n_r-k)!}
  \nonumber\\
  &\times&
  \frac{\Gamma(2\zeta+n_r+1)\Gamma(2\zeta+i+k)}{\Gamma(2\zeta+i+1) \Gamma(2\zeta+k+1)}
  \,
  \frac{1}{\frac{Z\alpha}{\eta}-\nu}
  \nonumber\\
  &\times&
  \Biggl\{
  \left[
    \left( \frac{Z\alpha}{\eta}-\nu \right)^2
    + (n_r-i)(n_r-k)
  \right]
  \nonumber\\
  &&
  - \frac{E_C(nl_j)}{m} \left( \frac{Z\alpha}{\eta}-\nu \right)
  (2n_r-i-k)
  \Biggr\}
  \,.
\end{eqnarray}

The base integrals, defined as \cite{rel1,rel2}
\begin{eqnarray}
  K_{abc}(\kappa)
  &=&
  \int_{0}^{1}{dv}\,\frac{v^{2a}}{(1-v^2)^{b/2}}\,
  \left(\frac{\kappa\sqrt{1-v^2}}{1+\kappa\sqrt{1-v^2}}\right)^{c}
  \nonumber\\
  &=&
  \int_0^1 y^{1-b} (1-y^2)^{a-1/2} \left( \frac{\kappa y}{1+\kappa y} \right)^{c}
  \;,
\end{eqnarray}
\[
  K_{bc}(\kappa)=K_{1bc}(\kappa)-\frac13 K_{2bc}(\kappa)
  \,.
\]
It is easy to obtain for the first derivative of $K$
\begin{equation}
  \frac{\partial K_{bc}}{\partial\kappa}
  =
  \frac{c}{\kappa^2} \, K_{b+1,c+1}(\kappa)
  \,.
\end{equation}
The integrals $K$ can be also expressed in a closed form
\cite{rel1,rel2}
\[
  K_{abc}(\kappa)=
  \frac{\kappa^c}{2}\,
  B\left(a+\frac12,1-\frac{b}{2}+\frac{c}{2}\right)
\]
\[
  \times
  {_3F_2}\left(\frac{c}{2},\, \frac{c}{2}+\frac12,\, 1-\frac{b}{2}+\frac{c}{2} ;\;
  \frac12,\, a+\frac32-\frac{b}{2}+\frac{c}{2} ;\; \kappa^2\right)
\]
\begin{equation}\label{defKabc}
  -\frac{c\,\kappa^{c+1}}{2}\,
  B\left(a+\frac12, \frac32-\frac{b}{2}+\frac{c}{2}\right)
\end{equation}
\[
  \times
  {_3F_2}\left(\frac{c}{2}+1,\, \frac{c}{2}+\frac12,\, \frac32-\frac{b}{2}+\frac{c}{2};\;
  \frac32,\, a+2-\frac{b}{2}+\frac{c}{2};\; \kappa^2\right)
\;,
\]
where $B\bigl(\alpha,\beta\bigr)$ is the beta
function and ${_3F_2}\bigl(\alpha,\beta,\gamma;\;\delta,\epsilon;\;
z\bigr)$ stands for the generalized hypergeometric function (see,
e.g., \cite{gen3f2}).

The solution above is a solution of the Dirac equation for a
particle with mass $m$. However, as we see from Eq.~(\ref{eq:1:nr}),
the two-body energy is the easiest to express in terms of a Dirac
equation with the reduced mass, introducing corrections.

In muonic hydrogen for $n=2$ the argument of the hypergeometric
function, $\kappa'_2$, is less than unity ($\simeq 0.7$), and the
hypergeometric series converges well. For $n=1$ in muonic
hydrogen or $n=2$ in muonic helium, one has to use  analytic
continuation of the hypergeometric series or integral representation
of the hypergeometric function.


To calculate the term with derivative and the term with $\delta H$
we need efficient nonrelativistic expressions. The Uehling
correction in the nonrelativistic limit is
\begin{equation}
  f^{\rm (NR)}_U(\Za,\kappa)
  =
  \frac{\alpha\Zab^2}{\pi n^2} F_{nl}^{\rm (NR)}(\kappa_{n})
  \,,
\end{equation}
where
\begin{equation}
  F_{nl_j}^{\rm (NR)}(\kappa_n)
  =
  -\sum_{i,k=0}^{n-l-1} B_{ik}^{\rm (NR)} \, K_{2,2l+i+k+2}(\kappa_n) 
  \,,
\end{equation}
\begin{eqnarray}
  B_{ik}^{\rm (NR)}
  &=&
  \frac{(-1)^{i+k}(n-l-1)!}{i!(n-l-i-1)!k!(n-l-k-1)!} 
  \nonumber\\
  &\times&
  \frac{(n+l)!(2l+i+k+1)!}{(2l+i+1)!(2l+k+1)!}
\end{eqnarray}
can be expressed in terms of elementary functions. Alternative
expressions for the nonrelativistic correction can be found in
\cite{soto,rel2}.

In a particular case of the ground state the result has a simple
form \cite{uehl_cl}
\begin{eqnarray}\label{eu1s}
F_{10}^{\rm (NR)}(\kappa) &=& - \frac{1}{3}\,\biggl\{
-\frac{4+\kappa^2-2\,\kappa^4}{\kappa^3}\cdot A(\kappa)
\nonumber\\
 &+&
\frac{4+3\,\kappa^2}{\kappa^3}\cdot \frac{\pi}{2}
-\frac{12+11\,\kappa^2}{3\,\kappa^2} \biggr\}\;,
\end{eqnarray}
where
\begin{equation}
  A(\kappa)=\frac{\arccos(\kappa)}{\sqrt{1-\kappa^2}}=
  \frac{\ln\left(\kappa+\sqrt{\kappa^2-1}\right)}{\sqrt{\kappa^2-1}}\,.\nonumber\\
\end{equation}

The nonrelativistic kernels $K_{bc}$ have only integer subscripts
and that allows useful recurrance relations (see
\cite{rel1,uehl_an}). Applying them we arrive at \cite{uehl_an}
\begin{eqnarray}
  &F_{nl}^{\rm (NR)}&(\kappa_n) =
  \frac{(n+l)!}{(n-l-1)!(2n-1)!}
  \sum_{i=0}^{n-l-1}
  \frac{1}{(2l+i+1)!}
  \,
  \frac{1}{i!}
\nonumber\\&\times&
 \left( \frac{(n-l-1)!}{(n-l-i-1)!} \right)^2
  \left(\frac{1}{\kappa_n}\right)^{2(n-l-1-i)}
 \\ \nonumber&\times&
  \left( \kappa_n^2 \frac{\partial}{\partial\kappa_n} \right)^{2(n-l-i-1)}
  \,
  \kappa_n^{2(l+i+1)}
  \,
  \left( \frac{\partial}{\partial\kappa_n} \right)^{2(l+i)}\\ \nonumber
&&  \frac{F_{10}^{\rm (NR)}(\kappa_n)}{\kappa_n^2}
  \label{fnlf10}
  \,.
\end{eqnarray}

\section{The analytic result for the relativistic recoil Uehling correction}

Using the expression for the relativistic Uehling energy
Eq.~\eq{DeltaEU} in terms of base integrals $K_{bc}$ and their
various properties \cite{rel1,rel2,uehl_an,soto,quacl,3f2}, for the
third term of Eq.~\eq{eq:1:all} we arrive at
\begin{eqnarray}
  &&
  \frac{\alpha\Zab^2}{\pi n^2} \frac{m_R^2}{M}
  \left( f_C-1 \right)\nonumber\\
&&\quad\times
  \sum_{i,k=0}^{n_r} B_{ik} 
  \frac{2\zeta+i+k}{\kappa'_n}
  K_{2,2\zeta+i+k+1}(\kappa'_n)
  \,.
\end{eqnarray}


The last term of Eq.~(\ref{eq:1:all}) corresponds
to a matrix element of the additional Hamiltonian \eq{u:dh}.
For the relativistic wave functions it can be rewritten in the form
\begin{eqnarray}
&-&\langle \psi |  \Bigl(
 \frac{V_U V_C}{M}
    + \frac{1}{4M} [V_C, [\mathbf{p}^2, W_U] ]
\nonumber\\
 && ~~ + \frac{1}{4M} [V_U, [\mathbf{p}^2, W_C] ]\Bigr) | \psi
 \rangle
\nonumber\\
&=& \frac{\alpha\Zab^2}{2M\pi}
  \int_0^1
  dv \, \rho_e(v) \langle\psi|
  \frac{\lambda e^{-\lambda r}}{r}
  |\psi\rangle
  \;,
\end{eqnarray}
which has the same structure of integration over $r$ as $f_U$  (cf.
Sec.~\ref{s:du}) and one can readily obtain for it an expression
which differs from one for the Uehling correction \eq{Fnlj} only by
a factor and the second indices of $K_{bc}$, arriving at
\begin{equation}\label{finalDE2}
  \frac{\alpha\Zab^4}{\pi} \frac{m_R^2}{M} \frac{\eta}{\Za n^2}
  \sum_{i,k=0}^{n_r} B_{ik} \frac{ K_{3,2\zeta+i+k}(\kappa'_n)}{\kappa'_n} 
  \,,
\end{equation}
or, for the nonrelativistic case
\begin{equation}\label{finalDE2NR}
  \frac{\alpha\Zab^4}{\pi} \frac{m_R^2}{M} \frac{1}{n^3}
  \sum_{i,k=0}^{n_r}
  B_{ik}^{\rm (NR)} \frac{K_{3,2l+i+k+2}(\kappa_n)}{\kappa_n} 
  \,.
\end{equation}


Combining the different parts of the expression \eq{eq:1:all}, we obtain for the
correction to the first order of $\alpha$
\[
  E_U = m_R f_U(\Za,\kappa) + \Delta E_U
  \,,
\]
where its relativistic recoil part is
\begin{eqnarray}\label{erelrec}
  \Delta E_U
  &=&
  \frac{\alpha\Zab^2}{\pi n^2} \frac{m_R^2}{M}
  \sum_{i,k=0}^{n_r} B_{ik}\nonumber\\ 
&\times&
  \Biggl[
    \left( f_C-1 \right)
    \biggl(
      K_{2,2\zeta+i+k}(\kappa'_n)\nonumber\\
&+&
      \frac{2\zeta+i+k}{\kappa'_n}
      K_{2,2\zeta+i+k+1}(\kappa'_n)
    \biggr)\nonumber\\
&+&
    \frac{\Za\eta}{\kappa'_n} K_{3,2\zeta+i+k}(\kappa'_n)
  \Biggr]
  \,.
\end{eqnarray}
Neglecting higher-order relativistic corrections, we find in order
$\alpha\Zab^4 m^2/M$
\begin{eqnarray}\label{deu_nr}
  \Delta E_U^{\rm (NR)}
  &=&
  \frac{\alpha\Zab^4}{\pi n^3} \frac{m_R^2}{M}
  \sum_{i,k=0}^{n-l-1}
  B_{ik}^{\rm (NR)}\nonumber\\&\times& 
  \Biggl[
    -\frac{1}{2n} K_{2,2l+i+k+2}(\kappa_n)\nonumber\\
    &&\phantom{9}
    -\frac{2l+i+k+2}{2n\kappa_n} K_{3,2l+i+k+3}(\kappa_n)\nonumber\\
    &&\phantom{9}
    +\frac{1}{\kappa_n} K_{3,2l+i+k+2}(\kappa_n)
  \Biggr]
  \,.
\end{eqnarray}

\section{Results for particular states}

There are several classes of states of interest. In this section we
consider two of them, namely circular states ($l=n-1$) and low lying
states ($n=1,2$). The former are quite insensitive to the nuclear
structure, allowing accurate {\em ab initio\/} calculations, and may
be of a ``metrological'' interest \cite{mumass,pimass}, while the
latter are the most sensitive to the nuclear structure and may be
applied to measure the nuclear charge radius \cite{nature,zumbro}.

Below we present results in closed form in terms of the generalized
hypergeometric function ${}_3F_2$ for arbitrary $Z$ and $\kappa_n$
and additionally the asymptotic behavior for large $\kappa_n$ is
investigated.

\subsection{Circular states}

For states with maximal orbital and angular momenta, i.e. $l=n-1$ and
$j=l+1/2$ there is no difference between $\kappa_n$ and
its relativistic analogue $\kappa'_n$, and $n_r=0$.
The expression \eq{erelrec} in this case can be transformed to
\begin{eqnarray}
  &&\Delta E_U
  =
  \frac{\alpha\Zab^2 }{n\zeta\pi}\,
  \frac{m_R^2}{M}\nonumber\\
  &\times&
  \biggl[
  \left( f_C-1 \right) \left(
    K_{2,2\zeta}(\kappa_n)
    +
    \frac{2\zeta}{\kappa_n} K_{3,1+2\zeta}(\kappa_n)
  \right)\nonumber\\
  &&
  +\frac{\Zab^2}{n\kappa_n}
  K_{3,2\zeta}(\kappa_n)
  \biggr]
\end{eqnarray}
or, neglecting higher-order terms in $\Za$,
\begin{eqnarray}
  &&\Delta E_U^{\rm (NR)}
  =
  \frac{\alpha\Zab^4}{\pi n^4}\,
  \frac{m_R^2}{M}\,\nonumber\\
  &\times&
  \frac{1}{\kappa_n}
  \biggl[
  -\frac{\kappa_n}{2}K_{2,2n}(\kappa_n)
  +n K_{3,2n}(\kappa_n)\nonumber\\
  &&
  -n K_{3,1+2n}(\kappa_n)
  \biggr]
  \,.
\end{eqnarray}

The asymptotics of the last expression for large $\kappa_n$ is
\begin{eqnarray}\label{circnge20}
  &&\Delta E_U^{\rm (NR)}
  =
  \frac{\alpha\Zab^4}{\pi n^3}
  \frac{m_R^2}{M}
  \nonumber\\
  &\times&
  \Biggl[
  -\frac{1}{3n} \left( \ln(2\kappa_n) -\psi(2n)+\psi(1)+\frac{1}{6} \right)
  \nonumber\\
  &&
  +\frac{2}{3(2n-1)}
  -\frac{\pi}{4\kappa_n}
  \nonumber\\
  &&
  +\frac{2n-1}{4\kappa_n^2}
  - \pi \frac{(n-2)(2n+1)}{18\kappa_n^3}
  +{\cal O}\left( \frac{1}{\kappa^4} \right)
 \Biggr]\;,
\end{eqnarray}
where $\psi(z)=\Gamma'(z)/\Gamma(z)$.

\subsection{The low-lying states}

The above results can be applied to the case of the $1s_{1/2}$ and
$2p_{3/2}$ states. In particular, for the nonrelativistic case
\begin{eqnarray}\label{low1}
  &&\Delta E_U^{\rm (NR)}(1s)
  =
  \frac{\alpha\Zab^4}{\pi}\,
  \frac{m_R^2}{M}\times
  \nonumber\\
  &&
  \frac{1}{36\kappa^3(\kappa^2-1)}
  \biggl[
  -6(2\kappa^6-3\kappa^4-12\kappa^2+10)A(\kappa)
  \nonumber\\
  &&\phantom{\frac{1}{36\kappa_n^3(\kappa^2-1)}[}
  +2\kappa(5\kappa^4+16\kappa^2-30)
  \nonumber\\
  &&\phantom{\frac{1}{36\kappa_n^3(\kappa^2-1)}[}
  -3\pi(3\kappa^4+7\kappa^2-10)
  \biggr]
  \,,
\end{eqnarray}
or, for large $\kappa$,
\begin{eqnarray}
  &&\Delta E_U^{\rm (NR)}(1s)
  =
  \frac{\alpha\Zab^4}{\pi}\,
  \frac{m_R^2}{M}\times
  \nonumber\\
  &&
  \quad\biggl[
  -\frac13 \ln(2\kappa)
  +\frac{17}{18}
  -\frac{\pi}{4\kappa}
  +\frac{1}{4\kappa^2}
  \nonumber\\
  &&
  \quad
  +\frac{\pi}{6\kappa^3}
  +{\cal O}\left( \frac{1}{\kappa^4} \right)
  \biggr]
  \,,
\end{eqnarray}

Other particular cases of interest are $2s_{1/2}$ and $2p_{1/2}$
states. For the nonrelativistic case relations for these states
can be written in a unified form%
\footnote{To come to this form from Eq.~\eq{deu_nr} one can use the
relation
\begin{equation} \label{c}
  K_{b,c} = \frac{1}{\kappa} K_{b+1,c+1} + K_{b,c+1}
  \,.
\end{equation}}
\begin{eqnarray}\label{low2}
  &&\Delta E_U^{\rm(NR)}(2l)
  =
  \frac{\alpha\Zab^4}{\pi} \frac{m_R^2}{32M}
  \nonumber\\
  &\times&
 \Biggl\{
   - K_{24}(\kappa_2)
   +\frac4{\kappa_2} \Bigl[ K_{34}(\kappa_2)
   -   K_{35}(\kappa_2) \Bigr]
   \nonumber\\
   &&
   +\frac{2(1-l)}{\kappa_2^3}
   \Bigl[
   \kappa_2 K_{44}(\kappa_2)
   \nonumber\\
   &&
   \phantom{+\frac{1-\nu}{\kappa_2^3}}
   + 4K_{54}(\kappa_2)
   -4K_{55}(\kappa_2)
   \Bigr]
 \Biggr\}
 \,,
\end{eqnarray}
or, for large $\kappa$,
\begin{eqnarray}
  &&\Delta E_U^{\rm(NR)}(2l)
  =
  \frac{\alpha\Zab^4}{\pi} \frac{m_R^2}{16M}
  \nonumber\\
  &\times&
  \Biggl\{
  \frac{1}{3} \left(
   -\ln (2 \kappa_2)
   + \frac{16-7l}{3}
  \right)
  -\frac{\pi}{2\kappa_2}
  \nonumber\\
  &&
  +\frac{l+2}{2}\frac{1}{\kappa_2^2}
  +\frac{2\pi(1-l)}{3}\frac{1}{\kappa_2^3}
  +\calO{\frac{1}{\kappa_2^4}}
  \Biggr\}
  \,.
\end{eqnarray}

The numerical results for the low-lying states in light muonic
atoms, obtained from (\ref{low1}) and (\ref{low2}) are summarized in
Table~\ref{t:low}.

\begin{table}[htbp]
 \begin{center}
 \begin{tabular}{|l|l|l|l|l|}
 \hline
Atom & $1s$ & $2s$ & $2p_{1/2}$ & $2p_{3/2}$  \\
 \hline
$\mu$H    &$0.182$ & $0.0381$ & $0.000901$ & $0.000901$\\
$\mu$D    &$0.180$ & $0.0388$ & $0.000968$ & $0.000968$\\
$\mu^3$He &$0.122$ & $0.0459$ & $0.00184$  & $0.00184$ \\
$\mu^4$He &$0.121$ & $0.0459$ & $0.00184$  & $0.00184$ \\
\hline
 \end{tabular}
\caption{Relativistic recoil eVP corrections for the low-lying
levels in muonic hydrogen. The units are
$\left({\alpha}/{\pi}\right)(Z\alpha)^4m_R^2/M$.\label{t:low}}
\end{center}
\end{table}

\section{Comparison with the Breit-type calculations}

Evaluations of relativistic recoil effects within the Grotch-type
approach developed in this paper and the standard Breit-type
technique (see, e.g., \cite{pach04,jentschura,pra}) are
complementary. Both produce for the eVP effects not only the leading
term of order $\alpha(Z\alpha)^4m^2/M$, but also certain higher-order
corrections. While the Grotch-type calculation provides us
with partial account of $\alpha(Z\alpha)^6m^2/M$ contributions, the
Breit-type calculation leads to an exact (in $m/M$) result for
the $\alpha(Z\alpha)^4m$ contribution.

The additional $\alpha(Z\alpha)^6m^2/M$ terms in the Grotch-type
approach are not so important as a simplification of pure recoil
contributions. The technique allows us to easily separate the leading
$\alpha(Z\alpha)^4m^2/M$ term from the higher-order effects and
calculate it much more easily than by means of the Breit-type
evaluation.

Meanwhile, such an evaluation is completely separated from the
non-recoil relativistic term. On the contrary, the Breit-type evaluation
produces both recoil and non-recoil relativistic contributions
within the same calculations, which allows additional crosschecks.

In this section we describe a rearrangement of the Breit-type
evaluation which would allow a direct comparison between the
Grotch-type and Breit-type results.

In paper \cite{pra} we have calculated the relativistic recoil
correction in question for low-lying states in light muonic atoms by
both mentioned methods. In both cases we can expand the correction by
powers of $m/M$
\begin{eqnarray}\label{recoilsplit}
E_U&=&\frac{\alpha}{\pi}\,(Z\alpha)^4m_R\biggl[C_0
+C_1 \,\frac{m_R}{M}\nonumber\\
&~&+C_2 \,\left(\frac{m_R}{M}\right)^2+\dots \biggr] \,,
\end{eqnarray}
where $C_0$ corresponds to the known non-recoil Uehling correction
to the energy, $C_1$ is found by the Grotch method, and the Breit
method provides both $C_1$ and $C_2$. The coefficients $C_1$
calculated by the two methods agree. That is a unique
expansion and the $C_0$ and $C_1$ coefficients obtained by both
methods should be the same. The Grotch-type approach produces $C_0$
and $C_1$, but not $C_2$.

The conventional Breit-type calculation does not produce the result
in such a form which makes the direct comparison difficult.

The Breit approach is based on an unperturbed Hamiltonian
\[
H^{(0)}=\frac{{\bf p}^2}{2m_R} + V(r)
\]
and thus the wave function does not include the muon and nuclear
mass separately, but only in a combination in the form of the reduced mass
$\phi^{(0)}=\phi(r;m_R)$.

Meantime, the standard Breit equation (see, e.g., \cite{IV})
\begin{eqnarray} \label{VBr}
  V_{\rm  Br}
  &=&-\left(\frac{1}{m^3}+\frac{1}{M^3}\right)\frac{{\bf p}^4}{8}\nonumber\\
  &+&\frac{Z\alpha}{8}\left(\frac{1}{m^2}+\frac{1}{M^2}\right)4\pi \delta^3({\bf r})\nonumber\\
  &+&Z\alpha\left(\frac{1}{4m^2}+\frac{1}{2mM}\right)\frac{{\bf L}\cdot{\mbox{\boldmath $\sigma$}}}{r^3}
    +\frac{Z\alpha}{2mM}4\pi \delta^3({\bf r})\nonumber\\
  &+&\frac{Z\alpha}{2mM}\left[\frac{1}{r^3}{\bf L}^2-{\bf p}^2\frac{1}{r}-\frac{1}{r}{\bf p}^2\right]
 \end{eqnarray}
explicitly depends on the muon mass $m$ and the nuclear mass $M$. The
eVP addition to the Breit Hamiltonian is of the form \cite{pach04}
\begin{eqnarray} \label{Veitia22}
 V_{\rm  Br}^{\rm VP}
 &=&\left(\frac{1}{8m^2}+\frac{1}{8M^2}\right)\nabla^2 V_U\nonumber\\
 &+&\left(\frac{1}{4m^2}+\frac{1}{2mM}\right)\frac{V^{\prime}_U}{r}{\bf L}\cdot{\mbox{\boldmath $\sigma$}}\nonumber\\
 &+&\frac{1}{2mM}\nabla^2\left[V_U-\frac{1}{4}(rV_U)^\prime\right]\nonumber\\
 &+&\frac{1}{2mM}\left[\frac{V^{\prime}_U}{r}{\bf L}^2+\frac{{\bf p}^2}{2}(V_U-rV_U^{\prime})
\right.\nonumber\\
 &&\left. +(V_U-rV_U^{\prime})\frac{{\bf p}^2}{2}\right]
 \end{eqnarray}
As a result, the matrix element of $ V_{\rm  Br}$ and $V_{\rm
Br}^{\rm VP}$ over $\phi^{(0)}(r;m_R)$ depends on $m, M, m_R$ and is
not suited for presentation in the form of (\ref{recoilsplit}).

To adjust the Breit Hamiltonian to this form
one has to rewrite it as a function of $m_R$ and $M$, but not $m$. The
related corrections to the Hamiltonian take the form
\begin{eqnarray} \label{new:VBr}
  V_{\rm  Br}
  &=&-\left(\frac{1}{m_R^3}-\frac{3}{m_R^2M}\right)\frac{{\bf p}^4}{8}\nonumber\\
  &+&\frac{Z\alpha}{8}\left(\frac{1}{m_R^2}-\frac{2}{m_R M}\right)4\pi \delta^3({\bf r})\nonumber\\
  &+&\frac{Z\alpha}{4m_R^2}\frac{{\bf L}\cdot{\mbox{\boldmath $\sigma$}}}{r^3}
    +\frac{Z\alpha}{2m_R M}4\pi \delta^3({\bf r})\nonumber\\
  &+&\frac{Z\alpha}{2m_R M}\left[\frac{1}{r^3}{\bf L}^2-{\bf p}^2\frac{1}{r}-\frac{1}{r}{\bf
  p}^2\right]\;,
 \end{eqnarray}
and
\begin{eqnarray} \label{newVeitia22}
 V_{\rm  Br}^{\rm VP}({\bf r})
 &=&\left(\frac{1}{8m_R^2}-\frac{1}{4m_R M}\right)\nabla^2 V_U\nonumber\\
 &+&\frac{1}{4m_R^2}\frac{V^{\prime}_U}{r}{\bf L}\cdot{\mbox{\boldmath $\sigma$}}\nonumber\\
 &+&\frac{1}{2m_R M}\nabla^2\left[V_U-\frac{1}{4}(rV_U)^\prime\right]\nonumber\\
 &+&\frac{1}{2m_R M}\left[\frac{V^{\prime}_U}{r}{\bf L}^2+\frac{{\bf p}^2}{2}(V_U-rV_U^{\prime})
\right.\nonumber\\
 &&\left. +(V_U-rV_U^{\prime})\frac{{\bf p}^2}{2}\right]\,,
 \end{eqnarray}
and here all terms which contribute to order $(Z\alpha)^4m^3/M^2$
and $\alpha(Z\alpha)^4m^3/M^2$ are neglected.

The perturbations (\ref{new:VBr}) and (\ref{newVeitia22}) directly
lead to eVP results in (\ref{recoilsplit}). We realized such
a rearrangement in \cite{pra} and obtained results by the Breit-type
approach, which completely agree with our Grotch-type approach
within an uncertainty of numerical integration (cf. Table~\ref{t:low}).

Such a rearrangement is applicable not only in the first order in
$\alpha$. In particular, we applied it in \cite{a2Za4} to second
order in $\alpha$ and obtained in that work relativistic recoil corrections
(of the first order in $m/M$) consistent with our calculation of the
relativistic recoil by the Grotch-type method \cite{III}.

\section{Conclusions}

In this paper a method for a calculation of relativistic recoil
effects developed previously \cite{I} was applied perturbatively for
the one-loop electronic-vacuum-polarization corrections. With the
results of the Dirac problem with Coulomb+Uehling potential already
known, the evaluation of the additional recoil correction has dealt
only with nonrelativistic wave functions. It was performed in
closed analytic form in terms of the same base integrals as required
for the calculation of the Uehling correction itself (cf.
\cite{rel1,rel2,uehl_cl,uehl_an,soto}).

We also found asymptotics for high $Z\alpha m/(m_e n)$ for the
circular and low-lying states. The low-lying states, $1s, 2s, 2p$,
are of particular interest in light muonic atoms, because they
provide us with perhaps the best opportunity to determine the
nuclear charge radius.

Some time ago, certain results were obtained within the Breit-type
\cite{pach04,jentschura} and Grotch-type approaches \cite{borie11}.
They have been discussed in part by us  in \cite{pra}.

Both calculations contain not only the leading eVP relativistic
recoil term of order $\alpha(Z\alpha)^4m^2/M$, but also certain
higher-order corrections. Here we explain in details how to compare
those calculations. A modification of the effective Breit Hamiltonian
is described, which allows us to avoid any higher-order effects in
the Breit type approach. As a result \cite{pra}, we agree with
\cite{jentschura} and disagree with \cite{borie05,borie11} and
\cite{pach04}. A discrepancy with the former is due to an
inappropriate gauge used in that work, while the discrepancy with the
latter is most probably due to a numerical error there.

Here, we applied the technique based on the presentation of the results
in terms of base integrals. However, this is not necessary. If it
is desired, one can solve the related Dirac equation numerically. As
we mentioned for the relativistic recoil correction by itself even
the Dirac equation is not necessary. A nonrelativistic
Schr\'odinger equation with an appropriate potential and subsequent
nonrelativistic perturbation theory is sufficient.

The approach can be extended further and it has been extended. In a
subsequent paper \cite{III} it is successfully applied to the second-order
eVP relativistic recoil corrections to order
$\alpha^2(Z\alpha)^4m^2/M$.

\section*{Acknowledgments}

This work was supported in part by DFG (grant GZ: HA 1457/7-2), RFBR
(under grants \#\# 12-02-91341 and 12-02-31741) and Dynasty
foundation. A part of the work was done during a stay of VGI and EYK
at  the Max-Planck-institut f\"ur Quantenoptik, and they are
grateful to it for its warm hospitality.


\appendix

\end{document}